 \documentclass[prl,a4paper,
twocolumn,amsmath, amssymb, floatfix,
nofootinbib,wrapfloat byrevtex]{revtex4}

 \usepackage{bm}
 \usepackage{amsmath,mathrsfs,upgreek,graphicx,placeins,float}
 \usepackage{color}
 \usepackage[normalem]{ulem}

 \newcommand{\beq}[1]{\begin{equation}\label{#1}}
 \newcommand{\eeq}{\end{equation}}
 \newcommand{\bea}[1]{\begin{eqnarray}\label{#1}}
 \newcommand{\eea}{\end{eqnarray}}
  \newcommand{\bec}[1]{\begin{center}\label{#1}}
  \newcommand{\eec}{\end{center}} 
  \newcommand{\ssA}{{\scriptscriptstyle\!A}}
  \newcommand{\ssB}{{\scriptscriptstyle\!B}}
  \newcommand{\sstyl}{\scriptscriptstyle\!}

\begin{document}

 \title{Complementarity of Gravitation Collapse (II)\\ XOB and the Damping Gravitational Waveform as Evidences}
 \author{Ding-fang Zeng}
 \email{dfzeng@bjut.edu.cn, ORCID: https://orcid.org/my-orcid?orcid=0000-0001-5430-0015}
 \affiliation{Beijing University of Technology, Chaoyang, Bejing 100124, P.R. China
 }

 \begin{abstract}
This is the second paper of our working series on the complementarity of gravitational collapse. In this paper we prove that dynamics of XOB (an eXact One-Body Method) under the weak-field-low-speed expansion matches with the post-newtonian series to all orders for the conservative part of binary dynamics in general relativity. Using XOB and an inner-structure modulated quadrupole formula, we generate gravitational waveforms for the black hole binary merger process agreeing with numeric relativity to 99\% degree. Basing on this agreement, we argue that black holes in numerical relativities have inner structures we propose in the complementarity of gravitational collapse.
\end{abstract}
 \maketitle
  
{\em Introduction} To calculate the Gravitational Waveforms (GWforms) of binary merger processes is a crucial undertaking both for their detection and feature-interpreting-aimed source tracing. Since massive bodies curve spacetime in a non superpositional way, the two-body problem in General Relativity (GR) cannot be reduced to one-body one as in Newtonian mechanics. This makes the relative motion between the two bodies rather difficult to predict and the GWforms rather hard to calculate efficiently. To obtain necessary number of GWforms for matched filtering, the practical way is decomposing the whole process into three stages, treating each of them independently and joining neighboring ones phenomenologically.  Typically, the inspiral stage is treated by Post-Newtonian (PN) approximation \cite{Blanchet9501,PatiWill0007,blanchet0209,pnApproxLivingReview,EricPoinssonBigBook} and Effective One-Body (EOB) method \cite{eob1998,eob2000,eob2009,Damour0005}, the merger process is modeled by Numerical Relativity  (NR)\cite{NmRel0507,NmRel0511,NmRel0511103,turduken2006,turduken2007,turduken2008,NR2018,NR2019}, while the ring-down era is handeled by Black Hole Perturbation Theory (BHPT) \cite{BHPTwaveform1970,BHPTwaveform1971,BHPTwaveform1975,QNMcardoso0905} \cite{tidalLoveNumber2009Poisson,tidalLoveNumber2009Dmaour,tidalLoveNumber2015,tidalLoveNumber2021,tidalDeformationNumeric}. In true applications, this division-and-joining strategy helps detecting the gravitational wave directly. However, it hides key physics of the process behind technique details and observational data. Especially, if the merger initials are two point-like singularities of Schwarzschild BHs, then why and how they can become the circular-line-like singularity of a single Kerr BH?

\begin{figure}[t]
\includegraphics[totalheight=28mm]{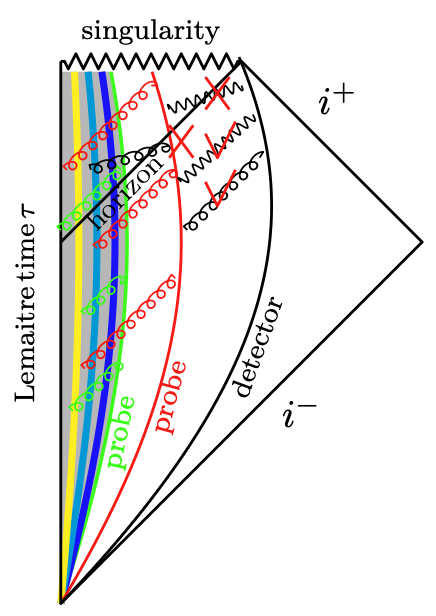}
\includegraphics[totalheight=28mm]{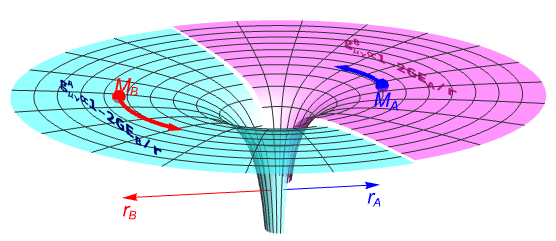}
\caption{Left, a self-gravitating collapsar indeed develops horizon and singularity in finite proper time. But its outside probe can only detect signals emitted before such things happen. So what the far away detector detects is only the history of the collapsar before its horizon and singularity appears. Right, XOB utilizes a double-patch-worked geometry as driving forces for the two merger bodies so that both of them move as test particles in static Schwarzschild geometries.}
\label{figRWmetric}
\end{figure}

BHPT can not address this question, because in it \cite{BHPTwaveform1970,BHPTwaveform1971,BHPTwaveform1975,QNMcardoso0905}, the ring-down feature arises from the in-falling and escaping conditions on background with axial symmetry already, e.g. Kerr-BHs.  NR defines the BH with their apparent horizon \cite{NmRel0507,NmRel0511,NmRel0511103} which lies inside the event horizon and whose shape deforms \cite{NR2018,NR2019} during the merger process. Although through the moving puncture \cite{turduken2006,turduken2007,turduken2008} or excision \cite{excise1965,excise1987,excise2002,excise2003} technique, NR showed that the data inside the apparent horizon affects not the GWform far away, using dynamics of the apparent horizon to determine the GWforms far away negates the event horizon wrapping singularity structure of BHs directly. So, there are two answers to the question why and how rotational symmetry is implemented during the merger process of BH binaries (BHB). The first is: the initial BHs are singular and the final rotational symmetry's implementation is a strong-gravitational effects non-describable in languages cultivated in our newtonian experience. The second is, from the beginning the BHs have regular internals and none horizons implied by CGC \cite{dfzeng2025a,dfzeng2023}, the final rotational symmetry's implementation is only a result of mass-transferring and re-distribution of the system. CGC does not contradict the singularity theorem. Because real BHs in both the NR's simulation and astronomical observations (AO) form through gravitational collapse, see the left Penrose-Carter diagram of FIG.\ref{figRWmetric} for reference. To all outside probe and detectors, they are asymptotical state with regular internals and no horizon in the Schwarzschild time definition rather than singular final state in the proper time definition. As long as their surface redshift is high enough, they are non-distinguishable from those described by the Schwarzschild or Kerr metric in isolation cases. To distinguish which of the two answers is right, a fully analytical approach to the BHB merger process or transparent computation of the corresponding GWform is almost the only way out. 

EOB method \cite{eob1998,eob2000,eob2009,Damour0005} applies not to the merger and ring-down phase of binary merger processes, during which the inner structure of BHs affect the GWform remarkably. The test particle executing the relative motion between the two BHs in this method falls into the horizon of the effective Schwarzschild-like background in finite coordinate time. This breaks the time dilation rule of GR directly and reflects an intrinsic inconsistence of this method. Ref.\cite{dfzeng2023} proposes XOB approach which applies to the whole three stages of the binary merger process uniformly. According to this approach, the GWform of binary merger process exhibits late time damping feature only when the merger participants carry no singularity or their point-like ``singularities'' elongate and bent so that the rotational symmetry enhances during the process. This is obviously an insight NR and PN would not tell us clearly and definitely. The problem is, ref.\cite{dfzeng2023} leaves the discrepancy between the weak-field-low-speed expansion of XOB and the conventional PN-approximation unresolved. This diminishes its persuasion remarkably. The current work will prove that, to all orders of PN-approximation, dynamics of the two methods are equivalent to each other. Differences between their serial hamiltonians are functions of their lower order conservative energy and angular momentum, so changes their zero-point references only. We will then calculate the GWform using XOB and an inner-structure modulated quadrupole formula, and compare with NR simulations, revealing physics related with the BH inner structure which NR alone would not tell us clearly and definitely.

{\em The key idea of XOB} is introducing a force-geometry consisting of two concentric patches $\{g^{\ssA}_{\mu\nu},g^{\ssB}_{\mu\nu}\}$ to support the orbital motion of the two bodies. Each patch accommodates a merger participant, but the two patches rotate synchronously so that each of them exert forces on its accommodations as a Schwarzschild BH does on its test particles, referring to FIG.\ref{figRWmetric}. The Lagrangian of the system reads, 
\bea{}
&&\hspace{-7mm}L(x^\mu_\ssA,x^\mu_\ssB){=}{-}M_{\ssA}\sqrt{{-}g^{\!B}_{\mu\nu}\dot{x}_A^{\mu}\dot{x}_A^{\nu}}{-}M_{\ssB}\sqrt{{-}g^A_{\mu\nu}\dot{x}_B^{\mu}\dot{x}_B^{\nu}}{+}L_\mathrm{diss} 
\label{GRtwobodyLagrange}
\\
&&\hspace{-7mm}g^{\ssB}_{\mu\nu}\dot{x}_{\ssA}^\mu\dot{x}_{\ssA}^\nu=
{-}h_{\ssB}\dot{t}^2_{\ssA}{+}h_{\ssB}^{-1}\dot{r}^2_{\ssA}
{+}r^2_{\ssA}\dot{\phi}^2_{\ssA},
h_{\ssB}{=}1{-}\frac{2GE_{\ssB}}{r_{\ssA}},
\label{metricBackgroundB}
\\
&&\hspace{-7mm}g^{\ssA}_{\mu\nu}\dot{x}_{\ssB}^\mu\dot{x}_{\ssB}^\nu=\mathrm{~similar~expressions},
\label{metricBackgroundA}
\eea
where $x^{\ssA}{=}\{t_{\ssA},r_{\ssA},\frac{\pi}{2},\phi_{\ssA}\}$ and $x^{\ssB}{=}\{t_{\ssB},r_{\ssB},\frac{\pi}{2},\phi_{\ssB}\}$ are world lines of the two bodies considered as point particles at this moment; $L_\mathrm{diss}$ is the dissipative part of the system when the GW radiation is considered.  Choosing affine parameters for the two particles independently $\dot{t}_\ssA=\dot{t}_\ssB=1$, their equations of motion read
\bea{}
&&\hspace{-5mm}\frac{h_\ssB}{\sqrt{h_\ssB{-}h_\ssB^{-1}\dot{r}_\ssA^2{-}r_\ssA^2\dot{\phi}_\ssA^2}}=\gamma_A,
\frac{\dot{\phi}_\ssA r_\ssA^2}{\sqrt{h_\ssB{-}h_\ssB^{-1}\dot{r}_\ssA^2{-}r_\ssA^2\dot{\phi}_\ssA^2}}=\ell_\ssA,
\label{EOMbinaryBody}
\\
&&\hspace{-5mm}\ddot{r}_\ssA{-}\frac{3h'_\ssB}{2h_\ssB}\dot{r}_\ssA^2
{+}(\dot{\phi}_\ssA^2-\frac{GE_\ssB}{r_\ssA^3})r_\ssA h_\ssB=0,
\label{EOMradiComponent}
\\
&&\hspace{-5mm}A\&B~\mathrm{exchanged~components}.
\label{EOMbinaryBodyB}
\eea  
$\gamma_{\ssA}$, $\ell_{\ssA}$ here are two conservative quantities. 
The parameter $E_{\ssB}$ in eqs\eqref{metricBackgroundB} and \eqref{EOMradiComponent} is an equivalent mass of the force geometry in which particle A moves. For particle B, similar conservative quantities or parameters $\gamma_\ssB,\ell_\ssB$ and $E_\ssB$ exist. The value of $E_\ssB\neq M_\ssB$, $E_\ssA\neq M_\ssA$ and $E_\ssA+E_\ssB<M_\ssA+M_\ssB$ due to the combination energy defect between the two bodies. 

For circular orbital motion, eq.\eqref{EOMradiComponent} and its $A\&B$ exchanged version yield
\beq{}
\dot{\phi}^2_\ssA=\frac{GE_\ssB}{r_\ssA^3},
\dot{\phi}^2_\ssB=\frac{GE_\ssA}{r_\ssB^3},
\label{Keppler3rdCircular}
\eeq
While for elliptic orbits with precession, eq\eqref{EOMbinaryBody} and its $A\&B$ exchanged version will tell us
\beq{}
\dot{\phi}_\ssA=\frac{\ell_\ssA}{\gamma_\ssA r_\ssA^2}\Big(1-\frac{2GE_\ssB}{r_\ssA}\Big),
\dot{\phi}_\ssB=\frac{\ell_\ssB}{\gamma_\ssB r_\ssB^2}\Big(1-\frac{2GE_\ssA}{r_\ssB}\Big).
\label{Keppler3rdElliptic} 
\eeq
If we define $\frac{\ell_\ssA}{\gamma_\ssA}\equiv\sqrt{GE_\ssB r_A}/(1{-}\frac{2GE_\ssB}{r_\ssA})$ and $\frac{\ell_B}{\gamma_B}$ similarly, eqs.\eqref{Keppler3rdElliptic} will reduce to \eqref{Keppler3rdCircular} routinely. For circular orbit motions, the central fixing and inspiral synchronizing conditions of the two bodies can be written and worked out directly
\bea{}
&&\hspace{8mm}m_\ssA r_\ssA-m_\ssB r_\ssB=0, 
\dot{\phi}_\ssA=\dot{\phi}_\ssB
\label{synchrConditionCircularA}
\\
&&\hspace{-5mm}r_{\ssA}=\frac{M_\ssB r}{M}, r_\ssB=\frac{M_\ssA r}{M}, E_\ssA=\frac{M_\ssA^3}{M^2}, 
E_\ssB=\frac{M_\ssB^3}{M^2}.
\label{synchrConditionCircularB}
\eea
where $M\equiv M_{\ssA}{+}M_{\ssB}$. For elliptic orbit motions with precession, eqs.\eqref{synchrConditionCircularB} cannot assure $\dot{\phi}_\ssA(t)=\dot{\phi}_\ssB(t)$ at all times. However, after their redshift factors $1{-}\frac{2GE_{\ssB(\ssA)}}{r_{\ssA(\ssB)}}$ are taken off, $\dot{\phi}_\ssA$ and $\dot{\phi}_\ssB$ equals to each other at all times. Considering this point, we identify eqs.\eqref{synchrConditionCircularB} as the central fixing and synchronicity conditions of the two bodies in general cases. Substituting them into eq\eqref{GRtwobodyLagrange}, the two-body Lagrangian $L(x^\mu_\ssA,x^\mu_\ssB)$ will become a single-body $L(r,\phi)$. Correspondingly, the Hamiltonian can be obtained by the standard Legendre procedure
\bea{}
H(r,\phi){=}\frac{M_\ssA(1-2GM^2_\ssB/rM)}{\sqrt{(1{-}\frac{2GM_\ssB^2}{rM}){+}(1{-}\frac{2GM_\ssB^2}{rM})^{-1}\frac{\dot{r}^2M_\ssB^2}{M^2}{-}\frac{r^2\dot{\phi}^2M_\ssB^2}{M^2}}}
\label{singlebodyAction}
\\
+\mathrm{A.B.exchanged.term}+L_\mathrm{diss}
\nonumber
\eea

Omitting the dissipation term temporarily, we will denote this Hamiltonian as $H^\mathrm{xob}$. Expanding $H^\mathrm{xob}$ in the weak-field and low-speed limit, defining $\mu\equiv\frac{M_\ssA M_\ssB}{M_\ssA{+}M_\ssB}$, $q\equiv\frac{r}{GM}$, we will have $H^\mathrm{xob}=\mu\sum c^{-2n}H_{2n}$, with the serial hamiltonian derivable routinely as 
\bea{}
H^\mathrm{xob}_0=\frac{1}{2}p^2{-}\frac{1}{q},
H^\mathrm{xob}_{-2}=\frac{1}{\nu}\sim(M_\ssA{+}M_\ssB)c^2,
\label{HamiltonianPN0}
\\
H^\mathrm{xob}_2=(1{-}3\nu)\big[\frac{3}{8} p^4{+}\frac{p^2{+}2(\bm{n}{\cdot}\bm{p})^2}{2 q}{-}\frac{1}{2 q^2}\big],
\label{HamiltonianPN1}
\\
H^\mathrm{xob}_4=(1{-}5\nu{+}5\nu^2)\big[\frac{5p^6}{16} +
\frac{9p^4{+}4p^2(\bm{n}{\cdot}\bm{p})^2}{8q}
\label{HamiltonianPN2}
\\
+\frac{3p^2+12(\bm{n}{\cdot}\bm{p})^2}{4 q^2}-\frac{1}{2q^3}\big],
\nu\equiv\frac{M_{\ssA}M_{\ssB}}{(M_{\ssA}{+}M_{\ssB})^2},
{\bm{n}}\equiv\frac{\bm{q}}{q}.
\nonumber
\eea
More higher serial hamiltonians can also be obtained easily. Comparing with the post-newtonian expressions of Poisson-Will \cite{EricPoinssonBigBook} and Damour-Shaffer  \cite{DamourPN1988},  $H^\mathrm{xob}_2$ and $H^\mathrm{xob}_4$ differ from $H^{\sstyl PW}_2$, $H^{\sstyl DS}_2$ and $H^{\sstyl DS}_4$. $H^{\sstyl PW}_2$ and $H^{\sstyl DS}_2$ differs from each other either. Ref.\cite{EricPoinssonBigBook} pedagogically explains the origin of non-uniqueness of PN Hamiltonians. Their differences arise from two reasons and can be killed correspondingly.  

The first is, XOB defines the radial coordinate $r$ in such a way that the spacetime metric has the perturbation order independent form of eq\eqref{metricBackgroundB}\&\eqref{metricBackgroundA} and rotates synchronously with the merger participants; while the PN-calculations of PW and DS define $r$ in such ways that their spacetime metrics exhibit general ADM 3+1 forms and get corrections order by order in $GM/r$. To compare different PN-hamiltonian series, we need transforming
\beq{}
q_{2n}^\mathrm{xob}\rightarrow q^{\sstyl PW,DS}(1+\frac{c_{1}^{2n}}{q}+\frac{c_{2}^{2n}}{q^2}+\cdots)^{\sstyl PW,DS}.
\label{qcoorRedefinition}
\eeq 
The second is, in obtaining the higher order corrections, PN's calculation used the conservation law of the lower order energy and angular-momentum,
\bea{}
H_{0}{+}\cdots c^{4{-}2n}H_{2n{-}4}{=}E_{2n{-}2}, q^2[p^2{-}({\bf n}{\cdot}{\bf p})^2]{=}L_{2n{-}2}^2.
\label{0pnCconservationLaw}
\eea
For any given $n$, through eqs.\eqref{qcoorRedefinition}-\eqref{0pnCconservationLaw}, we can write the differences $H^\mathrm{xob}_{2n}-H^{\sstyl PW,DS}_{2n}$ into zero or functions of $E_{2n{-}2}$ and $L_{2n{-}2}$, see the supplementary material for illustrations. Since $E_{2n{-}2}$ \& $L_{2n{-}2}$ are conservative relative to $H_{2n}$, the difference $H^\mathrm{xob}_{2n}-H^{\sstyl PW,DS}_{2n}$ becomes the difference of zero-point-references, their dynamics keep the same. Repeating this operation iteratively, we conclude that dynamics of XOB and PN are equivalent to each other to all orders. Schematically, their differences can be written as,
\bea{}
\mathrm{PN}:~(t,r)^\mathrm{metric}_{\sstyl ADM0}:>H_0:>(t,r)^\mathrm{metric}_{\sstyl ADM2}:>H_2:>\cdots,
\\
\mathrm{XOB}:~(t,r)_\mathrm{patch-worked}^\mathrm{rotating-geometry}:>H=\mu\sum_{n=0}^{\infty}c^{-2n}H_{2n},
\eea
where $:>$ means, in the left-hand-side coordinate system, we derive the right-hand-side hamiltonian. XOB obtains the all-order-hamiltonian with the cost of replacing the order-by-orderly PN-corrected true spacetime with an imagined, rotating and patch-worked geometry which carries no energy away.

Due to differences between the radial coordinate definitions, the weak-field-low-speed expanded hamiltonian of XOB cannot be used as checks for the calculation of PN hamiltonian. However, XOB brings us an exact and full-process applicable one-body description for the conservative part of the binary merger dynamics. Eqs.\eqref{EOMbinaryBody}-\eqref{EOMbinaryBodyB} and the synchronicity conditions \eqref{synchrConditionCircularB} can be solved formally following \cite{Darwin1959,Scott2004,Pound2011},
\bea{}
r(t){=}\frac{a(1{-}e^2)}{1{+}e\!\cos\!\chi},
\dot{\chi}^2=\dot{\phi}_\ssA^2\cdot[
1{-}\frac{2GM_\ssB^2(3{+}e\cos\chi)}{aM(1{-}e^2)}],
\label{orbitSolutionA}\\
\dot{\phi}^2_\ssA{=}\frac{\frac{GM(1{+}e\!\cos\!\chi)^4}{a^3(1{-}e^2)^3}
[1{-}\frac{2GM_\ssB^2(1{+}e\!\cos\!\chi)}{aM(1{-}e^2)}]^2}{1{-}\frac{4GM^2_\ssB}{aM(1-e^2)}{+}\frac{4G^2M^4_\ssB}{a^2M^2(1-e^2)}},
\label{orbitSolutionB}
\\
\mathrm{and~similar~expressions~with~} A\leftrightarrow B.~~
\nonumber
\eea 
When dissipations caused by the GW radiation are considered, $a$ and $e$ will become slowly-varying functions of $t$ relative to $\phi(t)$, while the energy and angular momentum flux balance conditions at infinity can be written as
\beq{}
\frac{dH(a,e)}{d[a,e]}\frac{d[a,e]}{dt}{=}{-}F_\mathrm{diss},
\frac{dA(a,e)}{d[a,e]}\frac{d[a,e]}{dt}{=}{-}J_\mathrm{diss},
\label{fluxBalanceCondition}
\eeq
The $H$ \& $A$'s expression in terms of $a$ \& $e$ for the arbitrary mass ratio of XOB can also be obtained following  \cite{Darwin1959}, see the supplementary material,
\bea{}
H=
\frac{M_{\ssA}\sqrt{1{-}\frac{4GM^2_\ssB}{aM(1-e^2)}{+}\frac{4G^2M^4_\ssB}{a^2M^2(1-e^2)}}}{
\sqrt{1-\frac{(3+e^2)GM_\ssB^2}{aM(1-e^2)}}
}{+}(A{\leftrightarrow}B),
\label{HbinaryEcc}
\\
A=\frac{GM_\ssA M_\ssB a(1-e^2)}{M\sqrt{\frac{aM(1-e^2)}{GM_B^2}}\sqrt{1{-}\frac{(3{+}e^2)GM_B^2}{aM(1-e^2)}}}
{+}(A{\leftrightarrow}B).
\label{JbinaryEcc}
\eea

Eqs.\eqref{fluxBalanceCondition} describe the periodically-averaged evolution instead of instantaneous dynamics of the orbital parameters. We will take the quadrupole radiation as the only contributor of dissipations. In the circular orbit case, $F^\mathrm{co}_\mathrm{diss}\sim\frac{32}{5}\frac{G\mu^2a^4\omega^6}{c^5}$, $J^\mathrm{co}_\mathrm{diss}\sim\frac{32}{5}\frac{G\mu^2a^4\omega^5}{c^5}$. When considering the eccentricity of the orbit \cite{EricPoinssonBigBook,Peters1963,Peters1964}, and the finite size of the merger bodies \cite{dfzeng2023}, the fluxes and the GWform will become
\bea{}
&&\hspace{-7mm}F_\mathrm{diss}{=}
F^\mathrm{co}_\mathrm{diss}\zeta^2\frac{1{+}\frac{73}{24}e^2{+}\frac{37}{96}e^4}{(1-e^2)^{7/2}},
\label{bnnPowerRadiation}
J_\mathrm{diss}{=}J^\mathrm{co}_\mathrm{diss}\zeta^2\frac{1{+}\frac{7}{8}e^2}{(1-e^2)^2}
\\
&&\hspace{-7mm}h_{\mathrm{+,\times}}^{\ell{m}{=}22}\propto\frac{2G^2\!M\mu }{aRc^4}\frac{\zeta H_{+,\times}}{1-e^2}, 
\zeta{=}\frac{\sin\!{4GMz/a}}{4GMz/a},
\label{gravitationalwaveformA}
\\
&&\hspace{-7mm}H_{+}{=}{{(e^2\!\cos\!2\phi{-}3e\cos\!\phi{-}2)\cos\!2\phi}\atop{~-\!e(2\sin\phi{+}e\sin\!2\phi)\sin\!2\phi}},
H_{\times}{=}{{\cdots\sin\!2\phi}\atop{+\!\cdots\cos\!2\phi}}\!,
\label{gravitationalwaveformAB}
\eea  
where $\zeta$ is a modulation factor determined by the shape-deformation and rotational-symmetry-enhancement feature of the system. For BHs with point-like singularity, such things would not happen so that $z=0$ or $\zeta\equiv1$. But for BHs with regular internals, in principle, we need to solve the Einstein equation with sources to find the form of $\zeta$. However, if we approximate the two BHs as two arcs stretching and bending,  ref.\cite{dfzeng2023} proves that it has the form displayed in \eqref{gravitationalwaveformA}. We will call this form of $\zeta$ as banana-shape-deformation factor, see FIG.\ref{figGWshape} in the next section for reasons.  Substituting eqs.\eqref{bnnPowerRadiation} into \eqref{fluxBalanceCondition} and combining with the $\dot{\phi}$ expression of \eqref{orbitSolutionA}, we will get a first order differential system for $a(t)$, $e(t)$ and $\phi(t)$ which can be solved numerically, after which eq.\eqref{gravitationalwaveformA} can be used to calculate the GWform directly.

\begin{figure}[t]
\includegraphics[totalheight=32mm]{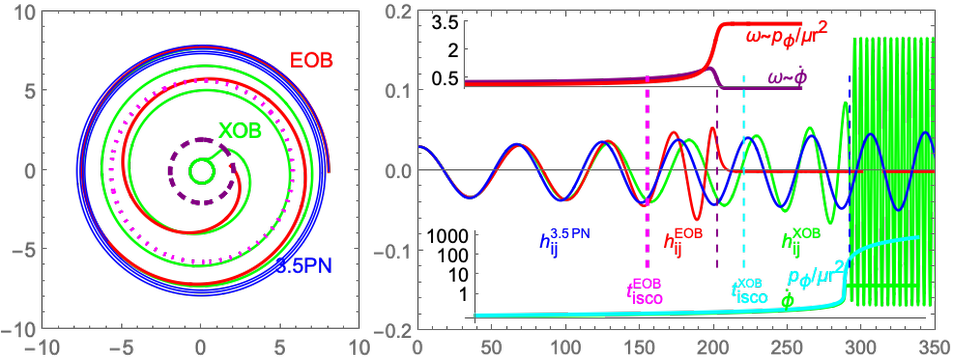}
\includegraphics[totalheight=25mm]{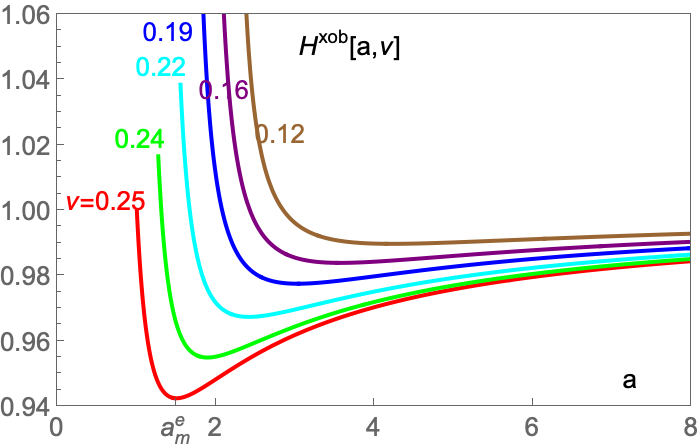}
\includegraphics[totalheight=25mm]{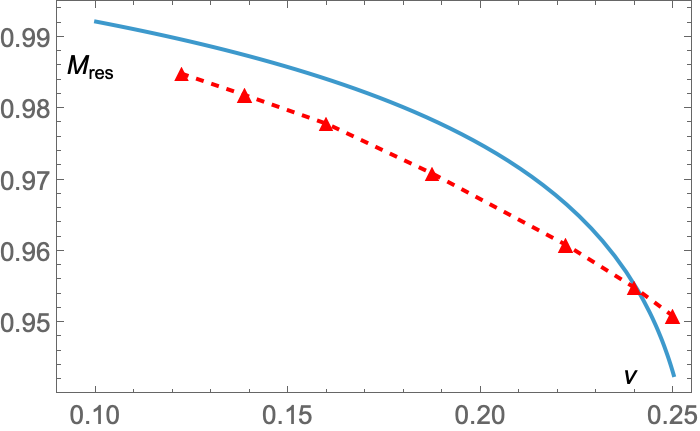}
\caption{The relative motion orbit and GWform of BHBs with central singularity and described by Schwarzschild metric exactly, upper panel. The red, green and blue curves are the prediction of EOB, XOB and 3.5PN approximation respectively. The lower-left is the separation dependence of binary system energy $H^\mathrm{xob}$ normalized to $M_\ssA{+}M_\ssB$; the lower-right is the residual mass of the merged system, the SXS simulation data is appended as red triangles.}
\label{fig3GW3compare}
\end{figure}

\begin{figure}[t]
\includegraphics[totalheight=40mm]{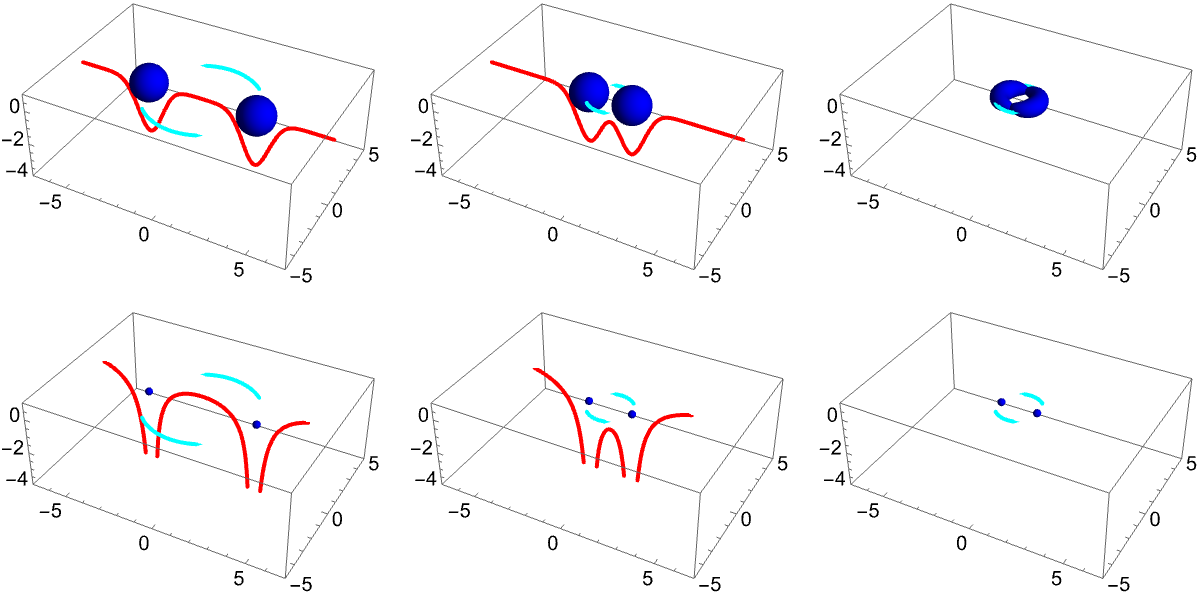}
\includegraphics[totalheight=32mm]{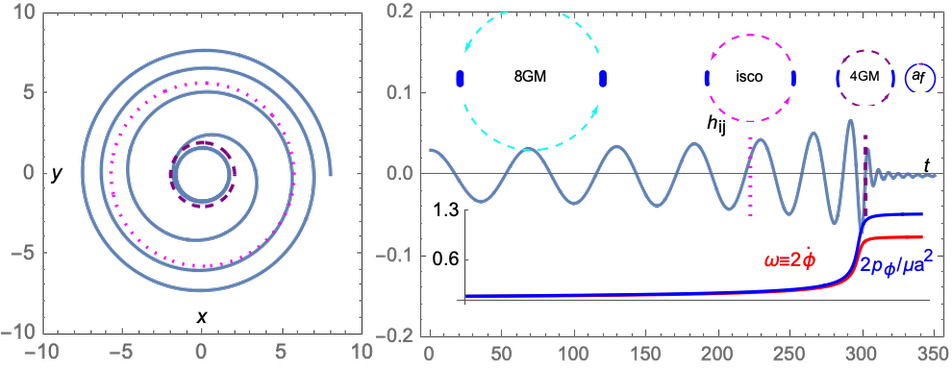}
\caption{Upper: when two black holes inspiral and get close, the potential barrier between them will decrease so that matter contents will transfer and cause rotational symmetry's strengthen if the black holes involved are regular internal and no-horizon-wrapped, but no similar things happen if the black holes are singular and horizon-wrapped. Lower: the relative motion orbit and GWform predicted by XOB for a binary system composed of BHs with regular internals and experience banana shape deformation under the inhomogeneous back reaction of GW radiation.}
\label{figGWshape}
\end{figure}

{\em GWforms of XOB and comparison} FIG.\ref{fig3GW3compare} displays the GWforms of three methods, EOB \cite{eob1998, eob2000}, XOB \cite{dfzeng2023} and the simple 3.5PN-approximation \cite{pnApproxLivingReview} directly, all for the merger of two BHs with point-like singularities. From the figure, we easily see that all three methods do not predict the late time feature of GWforms proplerly. The prediction of direct PN-approximation does not decay at all; that of EOB decays but to zero monotonically in a single cycle motion; that of XOB exhibits oscillation exotics but almost no amplitude decay. The PN's failure indicates that, this method misses key physics of the late time dynamics; the EOB's failure, especially its totally opposite trend of $\omega\equiv\frac{p_\phi}{\mu r^2}$ \& $\omega\equiv\dot{\phi}$ variation indicates that, this method contains intrinsic inconsistence. While XOB's failure points to, (i) decomposing the binary merger dynamics into conservative part and dissipation part is improper and, (ii) two BHs with point-like singularities cannot become rotationally symmetric in finite observational time.

We display in the lower-left part of FIG.\ref{fig3GW3compare} the separation dependence of the system energy. From the figure, we see that there is a minimal separation between the two BHs, on which $\frac{dH}{da}|_{a_m}=0$, and the energy of the system takes minimal value. $a_m$ is the stopping point of the differential system \eqref{fluxBalanceCondition} but not the terminal of the binary's inspiral motion. The no-decay of the late-time waveform is a result of decomposing the binary merger dynamics into Conservative part and Dissipative part. The C-part requires the system inspiral at frequency $\omega\sim GM/a_m^3$, eq.\eqref{orbitSolutionA}, the D-part allows it to radiate GWs according to \eqref{gravitationalwaveformA}. The waveform exotic magnifies the fact the system implements no rotational-symmetry from two initial BHs with point-like singularities. By identifying the minimum of $H^\mathrm{xob}$ as residual mass of the merger product, the lower-right of FIG.\ref{fig3GW3compare} compares the prediction of XOB and those of numeric simulation \cite{SXS2505}. The descrepancy still points to the fact that BHs of SXS are inner-structured. This will be analyzed further subsequently. The reason two BHs with point-like singularities will not get closer than $a_m$ or the energy of the system has a minimal at $a=a_m$ is a result of the time definition choice, see eqs.\eqref{metricBackgroundB}-\eqref{metricBackgroundA} and \eqref{singlebodyAction}. It is in this and other physically comparable time definitions, we see GWforms with ring-down feature from the astronomical observation and NR simulations. This fact is not treated properly by EOB because the test particle there \cite{eob1998, eob2000} executing the relative motion of the two BHs fall into the horizon $\sim2GM_\mathrm{tot}$ of the proxy background in a finite Schwarzschild time, so its GWform decays to zero monotonically in a semi-cycle oscillation, see the red waveform curve in the upper-right panel of FIG.\ref{fig3GW3compare} for references.

Practical BHs form through gravitational collapse. According CGC we propose in this working series, they are equally well described as collapsars with only close-to-implementing horizon in the Schwarzschild time definition and over-cross-oscillatory solid balls with periodically forming and resolving horizon and singularity. In the Schwarzschild time description, they may have very high surface redshift so that non-distinguishable in isolation cases from those described by exact Schwarzschild or Kerr metric \cite{dfzeng2025a,dfzeng2023,dfzeng2017,dfzeng2020,dfzeng2021,dfzeng2022,dfzeng2018a,dfzeng2018b,fuzzballsMathur2002,fuzzballSkenderis2007,mayerson2021fuzzball,mayerson2022microReview}. However, when two of them inspiral and get close enough, the potential barrier between them would decrease so that their matter contents can transfer from one to the other, see the upper part of Fig.\ref{figGWshape} for references. Combining with the inspiral motion, this radial mass transferring will change the mass configuration of the system from two spherically symmetric balls into a pair of comma-shaped irregular bodies \raisebox{-3pt}{`'}. The system will become more and more rotationally symmetric and radio-unwillingness. In eqs.\eqref{bnnPowerRadiation}-\eqref{gravitationalwaveformA}, we approximated this comma-pair shaped configuration with two concentric arcs and wrote it into the modulation of a banana-shape deformation factor $\zeta[z(t)]$, where $z(t)\equiv\frac{\ell_\mathrm{arc}(t)}{4GM}$ is the time dependent elongation multiplicity of the BHs when considered as two arcs of radius $a$. We will model phenomenologically
\beq{}
z(t)=z_f[\frac{4GMz_f}{\pi a(t)}\big]^{n_\mathrm{bnn}}.
\label{bnnZFitfunction}
\eeq 
At the same time, we will model the mass-transferring effect between the two BHs as follows
\beq{}
\mu{=}\frac{(M_\ssA{-}\Delta)(M_\ssB{+}\Delta)}{M},
\Delta{=}\frac{M_\ssA{-}M_\ssB}{2}e^{-\frac{k(a{-}a^\mathrm{e}_m)}{GM}},
\label{massTransferring}
\eeq
where $a^e_m=\frac{3}{2}GM$ is the minimal separation between the two BHs when their masses become equal, FIG.\ref{fig3GW3compare}, the lower-left part. The dimensionless parameters $z_f$, $n_\mathrm{bnn}$ and $k$ can be determined by fitting with NR-simulations. In principle, the value of these parameters should be universal.

\begin{figure}[t]
\includegraphics[totalheight=32mm]{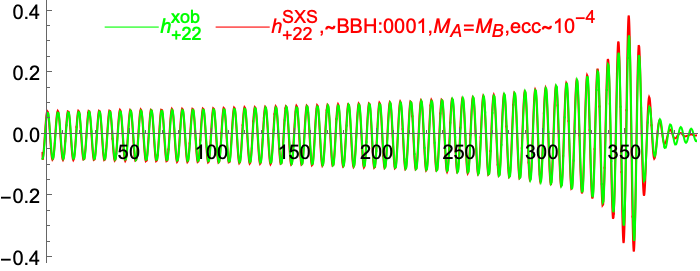}
\includegraphics[totalheight=32mm]{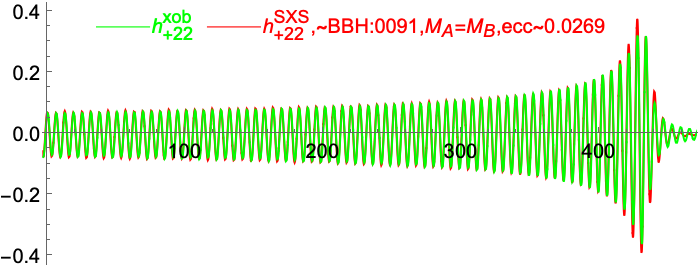}
\includegraphics[totalheight=32mm]{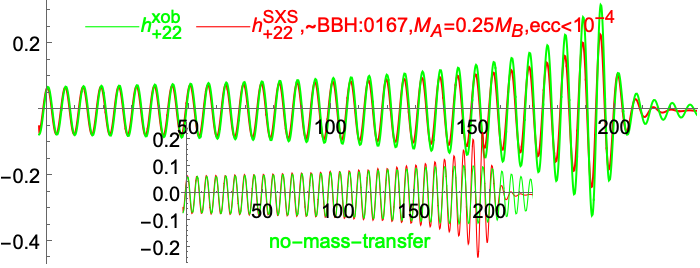}
\caption{Comparing the GWform of XOB and SXS numerical simulation BBH:0001,0091\&0167. Matching degrees of the three big comparisons are better than 99\%, 99\% \& 98\%. That of the bottom inset does not exceed 93\%.}
\label{figXobSxsComp}
\end{figure}

FIG.\ref{figXobSxsComp} compares the waveform of XOB and that of numeric relativity \cite{SXS2505} SXS:BBH:0001,0091,0167, all belongs to the type of no-spinning initial BHs. According to the definition of ref.\cite{mismatchLindblom2008},
\beq{}
\mathrm{mismatch}\equiv1-\frac{2\int h^\mathrm{xob}_{+22}(\phi)h^\mathrm{sxs}_{+22}(\phi)d\phi}{
\int|h^\mathrm{xob}_{+22}(\phi)|^2d\phi+\int|h^\mathrm{sxs}_{+22}(\phi)|^2d\phi},
\eeq
best fit yields
\bea{}
\begin{tabular}{ccccccc}
SXS.No.& sim.param&$a_0^\mathrm{xob}$&$z_f$&$n_\mathrm{bnn}$ & $k$ & mis
\\
\hline
BBH0001 & ${\frac{M_\ssA}{M_\ssB}{=}1,e{\sim}10^{-4}}\atop{a_0^\mathrm{sxs}{=}18}$&18.64 & 1.574 & 1.372&$-$ &0.0027
\\
BBH0091 & ${\frac{M_\ssA}{M_\ssB}{=}1,e{=}0.027}\atop{a_0^\mathrm{sxs}{=}19}$&20.13 & 1.659 & 1.618&$-$ &0.0030
\\
BBH0167 & ${\frac{M_\ssA}{M_\ssB}{=}\frac{1}{4},e{<}10^{-4}}\atop{a_0^\mathrm{sxs}{=}12.56}$&$13.23\atop{13.5}$ & $1.574\atop{4.162}$ & $3.271\atop{3.054}$&${0.306}\atop{0.112}$ & ${0.0192}\atop{0.078}$
\end{tabular}.
\nonumber
\eea
The value of $z_f$ exhibits degrees of universality, but that of $n_\mathrm{bnn}$ does not. This is because, the mass transferring \eqref{massTransferring} between the two BHs of BBH0167 affects the shape deformation progression \eqref{bnnZFitfunction}. This puts BBH0167 to different class of universality from BBH0001\&0091. The reason we choose comparing the GW stretching as functions of $\phi$ instead of $t$ is, the time definitions of XOB and NR are different. The time of XOB is defined through metrics \eqref{metricBackgroundB}\&\eqref{metricBackgroundA}; while that of NR is defined through the choice of lapse and shift functions in the ADM 3+1 metric ansatz. Since $h\sim{}e^{i2\phi(t)}$, $\dot{\phi}^2\sim GM/a^3$ and the radial coordinate used by XOB to measure the distance $a^\mathrm{xob}$ between the two BHs differs from that used by NR to measure $a^\mathrm{sxs}$, a function $a^\mathrm{xob}=r^\mathrm{xob}(a^\mathrm{sxs})$ similar with eq.\eqref{qcoorRedefinition} is needed to make the comparing of $h(t)$ functions in XOB and NR meaningful. More-later waveform or more-smaller-$a$-region physics is concerned, more severely is this relation needed to yield sensible matching score. Allowing $a_0^\mathrm{xob}$ be tunable and fitting it with numeric simulations, helps determining $a^\mathrm{xob}$'s global normalization relative to $a^\mathrm{sxs}$, but not its local dependence on $a^\mathrm{sxs}$ as $a^\mathrm{sxs}$ varies.  In principle, by tuning the lapse and shift function \cite{NmRel0507,NmRel0511,NmRel0511103}, the radial coordinate of NR can be aligned with that of XOB.

The mass would not be exchanged between two BHs with successfully implemented event horizon. However, for BHs with only infinitely-close to implementing but not truly implemented horizon, mass-transferring from the bigger and fatter one to the smaller and leaner one is natural due to the locally stronger attraction of the latter when the surfaces of two mergers touch each other. The inset of the bottom line of FIG.\ref{figXobSxsComp} displays, if one switches off the mass-transferring channel, then the matching degree between the XOB and SXS waveforms would not exceed 93\%. This implies that, between the two BHs of SXS:BBH:0167, mass transferring happens indeed. It is just the negligence of this mass-transferring that causes the descrepancy especially the crossing between the XOB predicted residual mass curve and that of NR  in the lower-right part of FIG.\ref{fig3GW3compare}. In NR, the relation between the feature of GWforms and the inner-structure of BHs is concealed behind the boundary data, including the in-falling conditions on the apparent horizon and the motion + shape-deformation of the apparent-horizon itself  \cite{turduken2006,turduken2007,turduken2008,NR2018,NR2019}.  
For lacking of reference from other independent computation, investigations of this relation are sidestepped or even refuted by the GW and BH physics community, although the observational data \cite{GWTC1,GWTC2,GWTC3} from the beginning suggests the ring-down and short-term-rising before the ring-down phenomenon's existence \cite{bhSpectroscopy2018,qnmObservation2021will,cardoso2016}.  As an fully analytical and independent computation with respect to NR-simulation, XOB reveals this relations transparently and definitely. The backward one body method of \cite{mcwilliams2019} has no this power because the damping feature of GWforms there is an input rather than a phenomena of rotational symmetry's enhancement due to the motion and re-distribution of matter sources.

{\em Conclusion} Through the radial coordinate transformation and conservation laws of the energy and angular-momentum iteratively, we prove in this work dynamics of the weak-field-low-speed expansion of XOB and that of conventional post-Newtonian approximation are equivalent to each other at all orders.  Using XOB and a banana-shape deformation factor modulated quadrupole formula for the radiation back reaction forces, we calculate the gravitational waveform of black hole binary merger processes with orbital but no spinning angular momentum. We get results matching the simulation of numerical relativity better than 99\%. As a fully analytical and whole-process-applicable one body formalism, XOB offers us not only a fast algorithm for the gravitational waveform template production, but also a powerful tool for their feature-interpreting-aimed provenance tracing. It tells us that, the late-time damping waveform of black hole binary merger process arises from their mass configuration's deformation and rotational symmetry enhancement. Two black holes with point-like singularities would not become a single Kerr hole with circular-line-like singularity; both the astronomical black holes and those simulated in the numerical relativity have inner-structures implied by the complementarity of gravitational collapse; they should be more considered as asymptotical state in the Schwarzschild time definition during which all outside probes are selected and detectors are monitored rather than singular final state defined in the proper time definition and detectable only to the co-moving probes.

XOB can be easily applied to the orbital motion of binary systems with non-zero eccentricity and arbitrary mass ratios. Its limitation is that, its time-domain waveform $h(t)$ is not directly comparable with numerical-relativity results because of the time coordinates in two approaches differ; the directly comparable quantity is the strain–phase relation $h(\phi)$. Its advantage includes the light computational working-load and the capability of linking the feature of gravitational waveforms to the inner-structure of BHs involved in the merger process. Such links are usually expressed as mosaics such as NR neither supports nor opposes the singularity's existence inside black holes.  XOB not only reveals such links clearly and definitely, but also derives quantitative predictions basing on them, which fit with numeric relativity and observations well \cite{dfzeng2023,GWTC1,GWTC2,GWTC3}.  In this work, we considered the merger of initially spherically symmetric black holes only. However no fundamental barrier exists prohibiting this method's applicability to the spinning black holes, after appropriate adaptations. This is an interesting direction for future works. 

{\bfseries Acknowledgements} We thank very much to professor Tom Osburn, Nick Houston for thoroughly reading and revision comments of our manuscript. Part of this work was completed at Niels Bohr Institute, with support from the Chinese Scholarship Council for oversea visiting research under grant no. CSC202006545026, and warmly hosted by Vitor Cardoso and Julie de Molade at the NBI. 
This work is also supported by the NSFC grant no. {11875082}, 
the Villum Investigator program of VILLUM Foundation (grant no. VIL37766) and the DNRF Chair program (grant no. DNRF162) by the Danish National Research Foundation and the European Union's Horizon 2020 research and innovation programme under the Marie Sklodowska-Curie grant No. 101131233.

\section*{Supplementary Material}

This supplementary material will address two questions. (A) provide detailed transformations $q^\mathrm{xob}_{2n}\rightarrow q^{\sstyl PW,DS}$ through which $H^\mathrm{xob}_{2n}-H^{\sstyl PW,DS}_{2n}$  becomes zero or functions of the lower order conservative quantities $E_{2n{-}2}$ and $L_{2n{-}2}$ iteratively. (B) Derive the hamiltonian, angular-momentum expression \eqref{HbinaryEcc}-\eqref{JbinaryEcc} of general relativistic test particles moving along processing orbit and their quadrupole formula \eqref{gravitationalwaveformA}, under the Darwin parameterization \eqref{orbitSolutionA}-\eqref{orbitSolutionB}.

{\it Part A} Let us begin from $H^\mathrm{xob}_{2}{-}H^{\sstyl DS}_{2}$, according to expressions \eqref{HamiltonianPN1} and those of the standard PN expansion \cite{pnApproxLivingReview,eob1998,DamourPN1988}, through the conservational laws
\beq{}
\frac{p^2}{2}-\frac{1}{q}=E_0(\mathrm{const}),q^2[p^2-(\mathbf{n}{\cdot}\mathbf{p})^2]=L_0^2(\mathrm{const}),
\label{consvLaw0pn}
\eeq 
and the transformation 
\bea{}
q_2^\mathrm{xob}=q^{\sstyl DS}(1{+}\frac{c_1^{(2)}}{q^{\sstyl DS}}{+}\frac{c_2^{(2)}}{q^{2\sstyl DS}}{+}\cdots),
\label{q2xobq0Transform}
\eea
we can write the difference between the XOB and Damour-Sch\"afer 1PN hamiltonian as
\bea{}
H^\mathrm{xob}_{2}{-}H^{\sstyl DS}_{2}
{=}(1-3 \nu )\big[2 E_0^2{-}
\frac{12 c_1^{(2)}E_0^2{+}\frac{19 \nu{-}10}{1{-}3 \nu}E_0}{q}{-}
\label{diff2xob}
\\
-\frac{\big(\frac{7{-}\frac{23 \nu }{2}}{3 \nu{-}1}{+}39 c_1^{(2)} E_0{-}36 c_1^{(2)2} E_0^2{+}18 c_2^{(2)} E_0^2\big)}{q^2}-\cdots
\nonumber
\\
\frac{\mathrm{polynomial}(\nu,\!E_0,\!L_0,\!c_1^{(2)}\!\cdots c_{k{-}1}^{(2)}){+}(6k{+}6)c_k^{(2)}E_0^2}{q^k}
{\cdots}\big]
\nonumber
\eea
Obviously, as long as we set
\bea{}
c_{1}^{(2)}{=}\frac{10{-}19 \nu}{12E_0(1{-}3 \nu)},\!
c_{2}^{(2)}{=}\frac{\frac{7{-}\frac{23 \nu }{2}}{3 \nu{-}1}{+}39c_1^{(2)}\!E_0{-}36 c_1^{(2)2}\!E_0^2}{18E_0^2},
\label{transXOBcoeff12}
\eea
\bea{}
\cdots,c_{k}^{(2)}{=}\frac{\mathrm{polynomial}(\nu,E_0,L_0,c_1^{(2)},c_2^{(2)}\cdots,c_{k{-}1}^{(2)})}{6(k{+}1)E_0^2}.
\label{transXOBcoeffk}
\eea
the difference $H^\mathrm{xob}_{2}{-}H^{\sstyl DS}_{2}$ becomes a conservative quantity $2 E_0^2(1{-}3 \nu )$ at this order as we claimed.

Now, we consider the difference $H_{4}^\mathrm{xob}-H_{4}^{\sstyl DS}$. Copying eq.\eqref{HamiltonianPN2} and corresponding expressions of $H^{\sstyl DS}_4$ from ref.\cite{pnApproxLivingReview,eob1998,DamourPN1988}, but this time using the transformation 
\bea{}
q_4^\mathrm{xob}=q^{\sstyl DS}(1{+}\frac{c_1^{(4)}}{q}{+}\frac{c_2^{(4)}}{q^{2}}{+}\cdots),
\eea 
and the 2nd order conservational laws
\bea{}
\frac{p^2}{2}-\frac{1}{q}+(1{-}3\nu)\big[\frac{3}{8} p^4{+}\frac{p^2{+}2(\bm{n}{\cdot}\bm{p})^2}{2 q}{-}\frac{1}{2 q^2}\big]=E_2,
\label{H2conservation}
\\
q^2[p^2-(\mathbf{n}{\cdot}\mathbf{p})^2]=L_2^2, ~E_2\&L_2~\mathrm{are~constant}\rule{8mm}{0pt}
\label{L2conservation}
\eea
we will find that, as long as we set
\bea{}
c_1^{(4)}=
(3 \nu -1)\{3 E_2(3 \nu -1)[2 F_2(8-35 \nu +11 \nu ^2)
\\
 -36+165 \nu-93 \nu ^2]
-(F_2-1) (20+-95 \nu+71 \nu ^2)
\nonumber\\
-18 E_2^2 (4{-}25 \nu{+}49 \nu ^2) (1{-}3 \nu )^2\}
/\{20 (1{-}5 \nu{+}5 \nu ^2)
\nonumber\\
\big[6 E_2(3 \nu{-}1){-}1\big]
\big[3 (F_2{-}3) E_2 (3 \nu{-}1){-}2 F_2{+}2)\big]\}
\nonumber
\eea
\bea{}
c_2^{(4)}=\frac{(1-3 \nu )^2}{20}\big\{6 E_2(3 \nu -1) \big[(199 F_2{-}122)\nu^2~
\\
-3 (11 F_2{+}38)\nu{+}15 F_2{+}2\big]{-}(F_2{-}1)(199 \nu^2{-}33 \nu{+}15)
\nonumber\\
+36 E_2^2 (43 \nu ^2{+}27 \nu{+}7) (1-3 \nu )^2\big\}
\nonumber\\
/\{(1{-}5 \nu{+}5 \nu ^2)[6 E_2(3 \nu -1)-1]
\nonumber\\
\big[3 (3 F_2{-}5) E_2 (3 \nu{-}1){-}2 F_2{+}18 E_2^2 (1{-}3 \nu )^2{+}2\big]\}
+~
\\
\big\{(3\nu{-}1)\big(F_2 [E_2(3 \nu{-}1){+}13]{+}38 E_2 (3 \nu{-}1){-}13\big)\frac{9c_1^{(4)}}{10}
\nonumber\\
\big\}/
\big\{F_2[9 E_2(3\nu{-}1){-}2]{+}18 E_2^2 (1{-}3 \nu )^2{+}E_2 (15{-}45 \nu )
\nonumber\\
{+}2\big\}+\frac{10}{3}c_1^{(4)2}
,F_2\equiv\sqrt{1+6(1-3\nu)E_2}
\nonumber
\eea
\bea{}
c_{k}^{(4)}{=}\mathrm{RationalFunction}(\nu,E_2,L_2,c_1^{(4)},\cdots,c_{k{-}1}^{(4)})
\eea
the 2PN difference of $H^\mathrm{xob}_{4}$ and $H^{\sstyl DS}_{4}$ will become a constant depending on $E_2$\&$L_2$ only,
\bea{}
H^\mathrm{xob}_{4}{-}H^{\sstyl DS}_{4}
{=}\frac{4(1{-}5\nu{+}5 \nu ^2)[3E_2(F_2{-}3)(3\nu{-}1){-}2F_2{+}2]}{27 (3 \nu -1)^3}
\eea

The difference $H^\mathrm{xob}-H^{\sstyl PW}$ can be similarly treated following eqs.\eqref{consvLaw0pn}-\eqref{transXOBcoeffk}, but different from $H^\mathrm{xob}-H^{\sstyl DS}$, $H^\mathrm{xob}-H^{\sstyl PW}$ can be set to zero by appropriately tuning the q-coordinate transformation.

The difference $H^{\sstyl PW}-H^{\sstyl DS}$ can also be treated following eqs.\eqref{consvLaw0pn}-\eqref{transXOBcoeffk}, but $H^{\sstyl PW}-H^{\sstyl DS}$ can only be set to a constant depending on $E_0$ and $L_0$ instead of zero.

{\it Part B} Following Darwin's parameterization in ref.\cite{Darwin1959}, we formally write the processing eccentric orbit controlled by the differential eqs.\eqref{EOMbinaryBody}-\eqref{EOMradiComponent} as follows
\bea{}
r_{\ssA}=\frac{a_\ssA(1-e^2_\ssA)}{1+e_\ssA\cos\chi_\ssA},
\label{DarwinParameterization}
\eea
where $\chi_\ssA=\chi_\ssA(\tau)$ is not the orbital phase angle $\phi_\ssA$ but an intermediate variable to be determined in the following. The processing effect will be encoded in the dependence of $\chi_\ssA$ on $\phi_\ssA$. Under the proper time parameterization, eqs.\eqref{EOMbinaryBody} can be written as
\beq{}
\dot{\phi}_\ssA r^2_\ssA=J_\ssA, (1-\frac{2GE_\ssB}{r_\ssA})\dot{t}_\ssA=H_\ssA,
\eeq
where $H_\ssA$ and $J_\ssA$ represent the specific energy and angular momentum of particle A in the gravitational field of effective mass $E_B$.
While the four-velocity normalization condition implies
\beq{}
\dot{r}^2_\ssA=H_\ssA^2-(1-\frac{2GE_\ssB}{r_\ssA})(1+\frac{J^2_\ssA}{r_\ssA^2})
\eeq
The energy $H_\ssA$ and angular momentum $J_\ssA$ can be written as the functions of the orbit parameters $a$ and $e$ exclusively
\beq{}
H_\ssA^2{=}\frac{(\tilde{p}_\ssA{-}2)^2{-}4e^2_\ssA}{\tilde{p}_\ssA(\tilde{p}_\ssA{-}3{-}e^2_\ssA)}
,J_\ssA^2{=}\frac{G^2E^2_\ssB\tilde{p}_\ssA^2}{(\tilde{p}_\ssA{-}3{-}e^2_\ssA)}
,\tilde{p}_\ssA\equiv\frac{a_\ssA(1{-}e^2_\ssA)}{GE_\ssB}
\label{EJapeRelation}
\eeq
When the total energy and angular-momentum of the binary system are concerned, we only need to write
\beq{}
H=M_\ssA\sqrt{\frac{(\tilde{p}_\ssA{-}2)^2{-}4e^2_\ssA}{\tilde{p}_\ssA(\tilde{p}_\ssA{-}3{-}e^2_\ssA)}}
+M_\ssB\sqrt{\frac{(\tilde{p}_\ssB{-}2)^2{-}4e^2_\ssB}{\tilde{p}_\ssB(\tilde{p}_\ssB{-}3{-}e^2_\ssB)}},
\eeq
\beq{}
J=M_\ssA\sqrt{\frac{G^2E^2_\ssB\tilde{p}_\ssA^2}{(\tilde{p}_\ssA{-}3{-}e^2_\ssA)}}
+M_\ssB\sqrt{\frac{G^2E^2_\ssA\tilde{p}_\ssB^2}{(\tilde{p}_\ssB{-}3{-}e^2_\ssB)}}.
\eeq
Let $E_\ssB=\frac{M_\ssB^3}{M^2}$, $E_\ssA=\frac{M_\ssA^3}{M^2}$, $a_\ssA=\frac{aM_\ssB}{M}$ and $a_\ssB=\frac{aM_\ssA}{M}$, these two expressions will become eqs.\eqref{HbinaryEcc}-\eqref{JbinaryEcc} rountinely.

Now, let us derive the differential equations satisfied by $\phi_\ssA$ and $\chi_\ssA$. Use the derivative tricks $\frac{dx}{dt}=\frac{dx/d\tau}{dt/d\tau}$ and the Darwin parameterization \eqref{DarwinParameterization}\& \eqref{EJapeRelation}
\bea{}
&&\frac{d\phi_\ssA}{dt_\ssA}=\frac{d\phi_\ssA/d\tau}{dt_\ssA/d\tau}=\frac{J_\ssA/r^2_\ssA}{H_\ssA/(1-2GE_\ssB/r_\ssA)}
\eea,
\bea
&&\big(\frac{d\phi_\ssA}{dt_\ssA}\!)^2{=}\frac{\frac{G M (1+e_\ssA \cos\chi_\ssA)^4}{a^3(1-e_\ssA^2)^3}\big[1{-}\frac{2 G M_\ssB^2(1{+}e_\ssA\cos\chi_\ssA)}{aM(1{-}e_\ssA^2)}\big]^2}{1-\frac{4 G M_\ssB^2}{1-e^2}+\frac{4G^2 M_\ssB^4}{a^2M^2(1-e^2)}};
\eea
\bea{}
\frac{d\chi_\ssA}{dt_\ssA}=\frac{\frac{d\chi_\ssA}{dr_\ssA}\cdot\frac{dr_\ssA}{d\tau}}{dt_\ssA/d\tau}
=\frac{(1{+}e_\ssA\cos\chi_\ssA)^2}{a_\ssA(1{-}e_\ssA^2)e_\ssA\sin\chi_A}
\frac{dr_\ssA/d\tau}{dt_\ssA/d\tau},
\eea
\bea{}
\big(\frac{d\chi_\ssA}{dt_\ssA}\big)^2
{=}\frac{\frac{G M (1+e_\ssA \cos\chi_\ssA)^4}{a^3(1-e_\ssA^2)^3}\big[1{-}\frac{2 G M_\ssB^2(1{+}e_\ssA\cos\chi_\ssA)}{aM(1{-}e_\ssA^2)}\big]^2}{1-\frac{4 G M_\ssB^2}{1-e^2}+\frac{4G^2 M_\ssB^4}{a^2M^2(1-e^2)}}\\
\cdot\big[1{-}\frac{2 G M_\ssB^2(3{+}e_\ssA\cos\chi_\ssA)}{aM(1-e_\ssA^2)}\big].
\nonumber
\eea
This proves eqs.\eqref{orbitSolutionA}-\eqref{orbitSolutionB}

Finally, let us consider the proof of quadrupole formula with eccentricity \eqref{gravitationalwaveformAB}. Strictly speaking, this formula applies to the close orbit of Newtonian gravity only. However, for eccentric orbit with procession in general relativity, as long as $a(t)$ \& $e(t)$ are slow functions of $t$ relative to $\phi(t)$ \cite{Peters1963,Peters1964}, this formula would provide good approximations. During the merger process of  black hole binaries, $e\rightarrow0$, $a\rightarrow\mathrm{const}$ due to the angular dissipation, so this formula can also be applicable in principle. Let us start with the general quadrupole formula,
\bea{}
h_+(t)=\frac{G}{Rc^4}(\ddot{Q}_{xx}-\ddot{Q}_{yy}),
h_\times(t)=\frac{G}{Rc^4}\ddot{Q}_{xy}
\eea
using definitions $Q_{ij}=\mu(x_ix_j-\frac{1}{3}\delta_{ij}r^2)$, $x=r(t)\cos\phi(t)$, $y=r(t)\sin\phi(t)$ and the angular-momentum conservation law $\frac{d}{dt}\dot{\phi}r^2\propto r\ddot{\phi}+2\dot{r}\dot{\phi}=0$, we can prove that
\bea{}
h_+(t)=\frac{G\mu}{Rc^4}\big[(\dot{r}^2{-}\dot{\phi}^2r^2{-}\frac{GM}{r})\cos2\phi{-}2r\dot{r}\dot{\phi}\sin2\phi\big]
\label{hplusExpression}
\\
h_\times(t)=\frac{G\mu}{Rc^4}\big[(\dot{r}^2{-}\dot{\phi}^2r^2{-}\frac{GM}{r})\sin2\phi{+}2r\dot{r}\dot{\phi}\cos2\phi\big]
\label{htimesExpression}
\eea
By the close orbit approximation, but allowing $a$\&$e$ to vary slowly as $t$ grows so that
\bea{}
r=\frac{a(1-e^2)}{1+e\cos\phi},\dot{\phi}r^2=\sqrt{GMa(1-e^2)}
\label{ellipticOrbit}
\\
\dot{\phi}=\big[\frac{GM}{a^3(1-e^2)}\big]^\frac{1}{2}(1{+}e\cos\phi)^2
\label{EllipticOrbitDotphi}
\\
\dot{r}=\frac{a(1{-}e^2)e\sin\phi}{(1{+}e\cos\phi)^2}=[\frac{GM}{a}]^\frac{1}{2}\frac{e\sin\phi}{\sqrt{1-e^2}}
\label{EllipticOrbitDotr}
\eea
Substituting eqs.\eqref{ellipticOrbit},\eqref{EllipticOrbitDotphi} and \eqref{EllipticOrbitDotr} into eqs.\eqref{hplusExpression}-\eqref{htimesExpression}, we will get eq.\eqref{gravitationalwaveformAB} of the maintext.


\begin{thebibliography}{00}

\bibitem{Blanchet9501}
L. Blanchet, T. Damour, B. R. Iyer, C. M. Will, and A. G. Wiseman,
``Gravitational-Radiation Damping of Compact Binary Systems to Second Post-Newtonian Order'',
{\em Phys. Rev. Lett.} {\bf 74} (1995) 3515, arXiv: gr-qc/9501027,
\href{https://doi.org/10.1103/PhysRevLett.74.3515}{doi.org/10.1103/PhysRevLett.74.3515}

\bibitem{PatiWill0007}
M. E. Pati, C. M. Will,
``Post-Newtonian Gravitational Radiation and Equations of Motion via Direct Integration of the Relaxed Einstein Equations. I. Foundations'',
{\em Phys. Rev.} {\bf D62} (2000) 124015, arXiv: gr-qc/0007087,
\href{https://doi.org/10.1103/PhysRevD.62.124015}{doi.org/10.1103/PhysRevD.62.124015}.

\bibitem{blanchet0209}
L. Blanchet, and Iyer, B. R., 
“Third post-Newtonian dynamics of compact binaries: Equations of motion in the center-of-mass frame”, 
{\it Class. Quantum Grav.}, {\bf20} (2003), 755, arXiv: gr-qc/0209089,
\href{https://doi.org/10.1088/0264-9381/20/4/309}{doi.org/10.1088/0264-9381/20/4/309}

\bibitem{pnApproxLivingReview}
L. Blanchet,
``Gravitational Radiation from Post-Newtonian Sources and Inspiralling Compact Binaries'',
{\em Living Rev. Relativity} {\bf 17} (2014) 2, arXiv: 1310.1528,
\href{https://doi.org/10.12942/lrr-2014-2}{doi.org/10.12942/lrr-2014-2}

\bibitem{EricPoinssonBigBook}
E. Poisson, \& C. M. Will,
`` Gravity, Newtonian, Post-Newtonian, Relativistic'',
chapter 12,
Cambridge University Press, 2014,
{\bf ISBN:} 9781139507486,
\href{https://doi.org/10.1017/CBO9781139507486}{doi.org/10.1017/CBO9781139507486}

\bibitem{DamourPN1988}
T. Damour and G. Sch\"afer, 
``Higher-Order Relativistic Periastron Advances and Binary Pulsars'',
Nuovo Cimento 10, 123 1988,
\href{https://doi.org/10.1007/BF02828697}{doi.org/10.1007/BF02828697}.

\bibitem{eob1998}
A. Buonanno and T. Damour
``Effective one-body approach to general relativistic two-body dynamics'',
{\it Phys. Rev.} {\bf D59} (1999) 084006,
\href{https://doi.org/10.1 103/PhysRevD.59.084006}{DOI:10.1103/PhysRevD.59.084006}

\bibitem{eob2000}
A. Buonanno and T. Damour,
``Transition from inspiral to plunge in binary black hole coalescences''
{\it Phys. Rev.}{\bf D62} (2000) 064015,
\href{https://doi.org/10.1103/PhysRevD.62.064015}{DOI:10.1103/PhysRevD.62.064015}

\bibitem{eob2009}
T. Damour, B. R. Iyer, and Alessandro Nagar,
``Improved resummation of post-Newtonian multipolar waveforms from circularized compact binaries'',
{\it Phys. Rev.}{\bf D79} (2009) 064004,
\href{https://doi.org/10.1103/PhysRevD.79.064004}{DOI:10.1103/PhysRevD.79.064004}

\bibitem{Damour0005}
T. Damour, P. Jaranowski, G. Schaefer,
``On the determination of the last stable orbit for circular general relativistic binaries at the third post-Newtonian approximation'',
{\em Phys. Rev.} {\bf D62} (2000) 084011, arXiv: gr-qc/0005034,
\href{https://doi.org/10.1103/PhysRevD.62.084011}{doi.org/10.1103/PhysRevD.62.084011}

\bibitem{NmRel0507}
F. Pretorius,
``Evolution of Binary Black-Hole Spacetimes'',
{\em Phys. Rev. Lett.} {\bf95} (2004) 121101, arXiv: gr-qc/0507014,
\href{https://doi.org/10.1103/PhysRevLett.95.121101}{doi.org/10.1103/PhysRevLett.95.121101}

\bibitem{NmRel0511}
M. Campanelli, C. O. Lousto, P. Marronetti, and Y. Zlochower,
``Accurate Evolutions of Orbiting Black-Hole Binaries without Excision'',
{\em Phys. Rev. Lett.} {\bf96} (2006) 111101, arXiv: gr-qc/0511048,
\href{https://doi.org/10.1103/PhysRevLett.96.111101}{doi.org/10.1103/PhysRevLett.96.111101}.

\bibitem{NmRel0511103}
J. G. Baker, J. Centrella, D. Choi \& et al,
``Gravitational-Wave Extraction from an Inspiraling Configuration of Merging Black Holes'',
{\em Phys. Rev. Lett.} {\bf96} (2006) 111102, arXiv: gr-qc/0511103,
\href{https://doi.org/10.1103/PhysRevLett.96.111102}{doi.org/10.1103/PhysRevLett.96.111102}.

\bibitem{turduken2006}
M. Ansorg, B. Brugmann, W. Tichy, 
``Single-domain spectral method for black hole puncture data'', 
{\em Phys. Rev.} {\bf D70} (2004) 064011, arXiv: gr-qc/0404056
\href{https://doi.org/10.1103/PhysRevD.70.064011}{doi.org/10.1103/PhysRevD.70.064011}

\bibitem{turduken2007}
Z. B. Etienne, J. A. Faber, Y. T. Liu et al, 
``Filling the holes: Evolving excised binary black hole initial data with puncture techniques'', 
{\em Phys. Rev.} {\bf D76} (2007) 101503,
\href{https://doi.org/10.1103/PhysRevD.76.101503}{doi.org/10.1103/PhysRevD.76.101503}.

\bibitem{turduken2008}
J. D. Brown,
``Puncture evolution of Schwarzschild black holes'',
{\it Phys. Rev.} {\bf D77} (2008) 044018,
\href{https://doi.org/10.1103/PhysRevD.77.044018}{doi: 10.1103/PhysRevD.77.044018}.

\bibitem{NR2018}
V. Varma, M. A. Scheel,H. P. Pfeiffer, 
``Comparison of binary black hole initial data sets'', 
{\em Phys. Rev.} {\bf D98} (2018) 104011, arXiv: 1808.08228,
\href{https://doi.org/10.1103/PhysRevD.98.104011}{doi.org/10.1103/PhysRevD.98.104011}.

\bibitem{NR2019}
D. Pook-Kolb, O. Birnholtz, B. Krishnan, and E. Schnetter,
``Interior of a Binary Black Hole Merger'',
{\em Phys. Rev. Lett.} {\bfseries 123} (2019) 171102, arXiv:1903.05626.
\href{https://doi.org/10.1103/PhysRevLett.123.171102}{doi.org/10.1103/PhysRevLett.123.171102}

\bibitem{BHPTwaveform1970}
C. V. Vishveshwara,
``Scattering of Gravitational Radiation by a Schwarzschild Black-hole'',
{\em Nature} {\bf vol.227} (1970) 936–938,
\href{https://doi.org/10.1038/227936a0}{doi.org/10.1038/227936a0}.

\bibitem{BHPTwaveform1971}
P., William H.
Long Wave Trains of Gravitational Waves from a Vibrating Black Hole,
{\em Astrophysical Journal}  {\bf vol.170} (1971) L105,
\href{http://dx.doi.org/10.1086/180849}{dx.doi.org/10.1086/180849}

\bibitem{BHPTwaveform1975}
S. Chandrasekhar, S. Detweiler,
``The quasi-normal modes of the Schwarzschild black hole'',
{\em  Astrophys. J.} {\bf 170}  (1971) L105,
\href{https://doi.org/10.1098/rspa.1975.0112}{doi.org/10.1098/rspa.1975.0112}

\bibitem{QNMcardoso0905}
E. Berti, V. Cardoso, A. O. Starinets,
``Quasinormal modes of black holes and black branes'',
{\em Class. Quantum Grav.} {\bf 26} (2009) 163001, arXiv: 0905.2975,
\href{https://doi.org/10.1088/0264-9381/26/16/163001}{doi.org/10.1088/0264-9381/26/16/163001}

\bibitem{tidalLoveNumber2009Poisson}
T. Binnington and E. Poisson,
``Relativistic theory of tidal Love numbers'',
{\it Phys. Rev.} {\bf D80} (2009) 084018,
\href{https://doi.org/10.1103/PhysRevD.80.084018}{DOI:10.1103/PhysRevD.80.084018}

\bibitem{tidalLoveNumber2009Dmaour}
T. Damour and A. Nagar,
``Relativistic tidal properties of neutron stars'',
{\it Phys. Rev.} {\bf D80} (2009) 084035,
\href{https://doi.org/10.1103/PhysRevD.80.084035}{DOI:10.1103/PhysRevD.80.084035}

\bibitem{tidalLoveNumber2015}
N. Gürlebeck,
``No-Hair Theorem for BHs in Astrophysical Environments'',
{\it Phys. Rev. Lett.} {\bf 114} (2015) 151102,
\href{https://doi.org/10.1103/PhysRevLett.114.151102}{DOI:10.1103/PhysRevLett.114.151102}

\bibitem{tidalLoveNumber2021}
A. Le Tiec and M. Casals,
``Spinning BHs Fall in Love'',
{\it Phys. Rev. Lett}{\bf 126} (2021) 131102,
\href{https://doi.org/10.1103/PhysRevLett.126.131102}{DOI:10.1103/PhysRevLett.126.131102}
 
\bibitem{tidalDeformationNumeric}
A. Bohn, L. E. Kidder, and S. A. Teukolsky,
``Toroidal horizons in binary BH mergers''
{\it Phys. Rev.} {\bf D94} (2016) 064009,
\href{https://doi.org/10.1103/PhysRevD.94.064009}{10.1103/PhysRevD.94.064009}

\bibitem{excise1965},
C. W. Misner,
``Wormhole Initial Conditions''
{\it Phys. Rev.} {\bf 118} (1960) 1110,
\href{https://doi.org/10.1103/PhysRev.118.1110}{doi: 10.1103/PhysRev.118.1110}

\bibitem{excise1987}
J. Thornburg,
``Coordinates and boundary conditions for the general relativistic initial data problem'',
{\it Class. Quantum Grav.} {\bf 4} (1987) 1119,
\href{https://iopscience.iop.org/article/10.1088/0264-9381/4/5/013}{doi: 10.1088/0264-9381/4/5/013}

\bibitem{excise2002}
G. B. Cook, 
``Corotating and irrotational binary black holes in quasicircular orbits'',
{\it Phys. Rev.} {\bf D65} (2002)  084003,
\href{https://doi.org/10.1103/PhysRevD.65.084003}{doi: 10.1103/PhysRevD.65.084003}

\bibitem{excise2003},
H. P. Pfeiffer,  L. E. Kidder, M. A. Scheel, S A. Teukolsky,
``A multidomain spectral method for solving elliptic equations'',
{\it Comput. Phys. Commun} {\bf 152} (2003) 253,
\href{https://doi.org/10.1016/S0010-4655(02)00847-0}{doi: 10.1016/S0010-4655(02)00847-0}

\bibitem{dfzeng2025a}
Ding-fang Zeng,
``Inside Black Holes, Singularity or Complementarity?'',
\href{https://arxiv.org/abs/2505.14750}{2505.14750}.

\bibitem{dfzeng2023}
Ding-fang Zeng, 
``Microscopic State of BHs and an Exact One Body Method for Binary Dynamics in General Relativity'',
{\em EPJC} {\bf 84} (2024) 370,
\href{https://doi.org/10.48550/arXiv.2311.11764}{arXiv: 2311.11764}

\bibitem{Darwin1959}
C. Darwin,
``The gravity field of a particle'',
{\it Proc. Roy. Soc. Lond.}, {\bf A249} (1959) 180,
\href{https://doi.org/10.1098/rspa.1959.0015}{doi:10.1098/rspa.1959.0015}

\bibitem{Pound2011}
E. Poisson, A. Pound \& I. Vega, 
``The motion of point particles in curved spacetime'',
{\it Living Reviews in Relativity} {\bf 14} (2011) 7,
\href{https://doi.org/10.12942/lrr-2011-7}{doi: 10.12942/lrr-2011-7}

\bibitem{Scott2004}
S. Drasco and S. A. Hughes,
“Rotating black hole orbit functionals in the frequency domain.”
{\it Phys. Rev.} {\bf D69} (2004) 044015,
\href{https://doi.org/10.1103/PhysRevD.69.044015}{doi:10.1103/PhysRevD.69.044015}

\bibitem{Peters1963}
P. C. Peters and J. Mathews,
``Gravitational Radiation from Point Masses in a Keplerian Orbit'',
{\it Phys. Rev} {\bf 131} (1963) 435,
\href{https://doi.org/10.1103/PhysRev.131.435}{doi: 10.1103/PhysRev.131.435}.

\bibitem{Peters1964}
P. C. Peters,
``Gravitational Radiation and the Motion of Two Point Masses'',
{\it Phys. Rev} {\bf 136} (1964) B1224,
\href{https://doi.org/10.1103/PhysRev.136.B1224}{doi: 10.1103/PhysRev.136.B1224}.

\bibitem{SXS2505}
M. A. Scheel, M. Boyle K. Mitman and et al,
``The SXS Collaboration's third catalog of binary black hole simulations''
\href{https://doi.org/10.48550/arXiv.2505.13378}{doi: 10.48550/arXiv.2505.13378},
\href{https://arxiv.org/abs/2505.13378}{arxiv: 2505.13378}.

\bibitem{mismatchLindblom2008}
L. Lindblom, B. J. Owen, and D. A. Brown, 
``Model waveform accuracy standards for gravitational wave data analysis'',
{\it Phys. Rev.} {\bf D78} (2006), 124020,
\href{https://doi.org/10.1103/PhysRevD.78.124020}{doi: 10.1103/PhysRevD.78.124020}

\bibitem{dfzeng2017}
Ding-fang Zeng, 
``Resolving the Schwarzschild singularity in both classic and quantum gravities'',
{\em Nucl. Phys.} {\bf B917} (2017) 178, arXiv: 1606.06178,
\href{https://doi.org/10.1016/j.nuclphysb.2017.02.005}{doi.org/10.1016/j.nuclphysb.2017.02.005}.

\bibitem{dfzeng2020} 
Ding-fang Zeng, 
``Exact inner metric and microscopic state of AdS$_3$-Schwarzschild BHs'',
{\em Nucl. Phys.} {\bf B954} (2020) 115001, arXiv: 1812.06777,
\href{https://doi.org/10.1016/j.nuclphysb.2020.115001}{doi.org/10.1016/j.nuclphysb.2020.115001}.

\bibitem{dfzeng2021}
Ding-fang Zeng, 
``Spontaneous Radiation of BHs'',
{\em Nucl. Phys.} {\bf B977} (2022) 115722, arXiv: 2112.12531,
\href{https://doi.org/10.1016/j.nuclphysb.2022.115722}{doi.org/10.1016/j.nuclphysb.2022.115722}

\bibitem{dfzeng2022}
Ding-fang Zeng, 
``Gravity Induced Spontaneous Radiation'',
{\em Nucl. Phys.} {\bf B990} (2023) 116171,arXiv: 2207.05158,
\href{https://doi.org/10.1016/j.nuclphysb.2023.116171}{doi.org/10.1016/j.nuclphysb.2023.116171}

\bibitem{dfzeng2018a}
Ding-fang Zeng, 
``Schwarzschild Fuzzball and Explicitly Unitary Hawking Radiations'',
{\em Nucl. Phys.} {\bf B930} (2018) 533-544, arXiv: 1802.00675,
\href{https://doi.org/10.1016/j.nuclphysb.2018.03.012}{doi.org/10.1016/j.nuclphysb.2018.03.012}.

\bibitem{dfzeng2018b}
Ding-fang Zeng, 
``Information missing puzzle, where is hawking's error?'',
{\em Nucl. Phys.} {\bf B941} (2018) 665, arXiv: 1804.06726,
\href{https://doi.org/10.1016/j.nuclphysb.2019.02.023}{doi.org/10.1016/j.nuclphysb.2019.02.023}.

\bibitem{fuzzballsMathur2002}
O. Lunin and S. D. Mathur,
``Statistical Interpretation of the Bekenstein Entropy for Systems with a Stretched Horizon'',
{\it Phys. Rev. Lett.} {\bf 88}(2002) 211303,
\href{https://doi.org/10.1103/PhysRevLett.88.211303}{DOI:10.1103/PhysRevLett.88.211303}

\bibitem{fuzzballSkenderis2007}
K. Skenderis and M. Taylor,
``Fuzzball Solutions for BHs and D1-Brane–D5-Brane Microstates'',
{\it Phys. Rev. Lett.} {\bf 98} (2007) 071601,
\href{https://doi.org/10.1103/PhysRevLett.98.071601}{DOI:10.1103/PhysRevLett.98.071601}.

\bibitem{mayerson2021fuzzball}
F. Bacchini, D. R. Mayerson, B. Ripperda et al,
``Fuzzball Shadows: Emergent Horizons from Microstructure'',
{\it Phys. Rev. Lett.} {\bf127} (2021) 171601 

\bibitem{mayerson2022microReview}
D. R. Mayerson,
``Modave Lectures on Horizon-Size Microstructure, Fuzzballs and Observations'',
\href{https://arxiv.org/abs/2202.11394}{arxiv:2202.11394}

\bibitem{GWTC1}
``GWTC-1: A Gravitational-Wave Transient Catalog of Compact Binary Mergers Observed by LIGO and Virgo during the First and Second Observing Runs'', 
GWTC-1: {\em Phys. Rev.} {\bf X9} (2019) 031040, arXiv: 1811.12907,
\href{https://doi.org/10.1103/PhysRevX.9.031040}{doi.org/10.1103/PhysRevX.9.031040}.

\bibitem{GWTC2}
``GWTC-2: Compact Binary Coalescences Observed by LIGO and Virgo During the First Half of the Third Observing Run'',
GWTC-2: {\em Phys. Rev.} {\bf X11} (2021) 021053,
\href{https://doi.org/10.1103/PhysRevX.11.021053}{doi.org/10.1103/PhysRevX.11.021053}.

\bibitem{GWTC3}
``GWTC-3: Compact Binary Coalescences Observed by LIGO and Virgo During the Second Part of the Third Observing Run'',
GWTC-3: \href{https://doi.org/10.48550/arXiv.2111.03606}{doi.org/10.48550/arXiv.2111.03606}.

\bibitem{bhSpectroscopy2018}
V. Baibhav, E. Berti, V. Cardoso, and G. Khanna,
``BH spectroscopy: Systematic errors and ringdown energy estimates''
{\it Phys. Rev.} {\bf D97} (2018) 044048,
\href{https://doi.org/10.1103/PhysRevD.97.044048}{DOI:10.1103/PhysRevD.97.044048}

\bibitem{qnmObservation2021will}
M. Isi, W. M. Farr,
``Analyzing black-hole ringdowns''
\href{https://doi.org/10.48550/arXiv.2107.05609}{DOI:10.48550/arXiv.2107.05609}

\bibitem{cardoso2016}
V. Cardoso, S. Hopper, C F. B. Macedo and et al,
``Gravitational-wave signatures of exotic compact objects and of quantum corrections at the horizon scale'',
{\it Phys. Rev.} {\bf D94} (2016) no.8, 084031,
\href{https://arxiv.org/abs/1608.08637}{arXiv: 1608.08637}.

\bibitem{mcwilliams2019}
S. T. McWilliams,
``Analytical Black-Hole Binary Merger Waveforms'',
{\it Phys. Rev. Lett.} {\bf 122} (2019) 191102,
\href{https://doi.org/10.1103/PhysRevLett.122.191102}{doi:10.1103/PhysRevLett.122.191102}

\end{thebibliography}
\end{document}